\documentclass[preprint,12pt]{aastex}
\newcommand{\be}{\begin{equation}}
\newcommand{\ee}{\end{equation}}
\newcommand{\bea}{\begin{eqnarray}}
\newcommand{\eea}{\end{eqnarray}}
\newcommand{\rmscr}[1]{{\hbox{\scriptsize \rm{#1}}}}
\newcommand{\rmmat}[1]{{\hbox{\rm{#1}}}}
\begin{document}

\title{Rotational Evolution of Protoneutron Stars}

\author{Yefei Yuan \altaffilmark{1,2} and Jeremy S. Heyl\altaffilmark{2,3}}
\altaffiltext{1}{Center for Astrophysics, University of Science and Technology
of China, Hefei, Anhui 230026, P.R. China; yfyuan@ustc.edu.cn}
\altaffiltext{2}{Harvard-Smithsonian Center for Astrophysics, Cambridge, 
MA 02138; yyuan@cfa.harvard.edu; jheyl@cfa.harvard.edu}
\altaffiltext{3}{Chandra Fellow}
%\email{yyuan@cfa.harvard.edu}

\begin{abstract}
We study the rotational evolution of a protoneutron star with hyperons
and nucleons or solely nucleons in its core due to the escape of the
trapped neutrinos.  It is found that at the early stage of its
evolution, the stellar crust contracts significantly, consequently the
star spins up.  At the late stage, a the protoneutron star with
hyperons, it keeps shrinking and spinning up till all the trapped
neutrinos escape.  Consequently, the distribution of the stellar
initial spin periods is skewed toward shorter periods.  For a
protoneutron star with only nucleons, the expansion of its core
dominates, and the stellar rotation slows down.  After the neutrinos
escape, the range of the spin periods is narrower than the initial
one, but the distribution is still nearly uniform.  If the hyperonic
star is metastable, its rotational frequency accelerates
distinguishedly before it collapses to a black hole.
\end{abstract}

\keywords{dense matter --- stars: evolution, neutron, rotation}

\section{Introduction}
\label{sec:intr}

A neutron star (NS) is born after the iron core collapse of a massive
progenitor star ($>8M_{\sun}$) which exhausts its nuclear fuel.  The
birth of the neutron star often results in a Type-II supernova
explosion \citep{Burr00}.  Immediately after its birth, the neutron
star is hot and lepton-rich because the core is opaque to neutrinos.
This young object is called a protoneutron star (PNS).  As
compared to the neutrino-free case ({\em i.e.} the cold neutron star), the
trapped neutrinos significantly change the chemical equilibria between
baryons and leptons, the fractions of all the compositions, and the
equation of state (EOS).  Therefore, the
Kelvin-Helmholtz epoch of the evolution of the PNS, during which the
PNS changes from a hot and lepton-rich compact star to a cold and
neutrino-free one, is the most important evolutionary stage of the
PNS.  The time scale of the Kelvin-Helmholtz epoch completely depends
on the neutrino interactions in the hot and dense matter
\citep{Redd98}.  Namely the diffusion time scale of the trapped
neutrinos from the core to the outside is roughly several tens of seconds
\citep{Prak97}.  As pointed out in the literature, in principle, 
observations of supernova neutrinos may constrain the EOSs and the
interior composition of neutron stars.
\citep{Prak95,Pons99,Pons01prl,Pons01}.

The properties of PNSs which contain only ordinary nuclear matter
have been investigated in detail by many authors \citep{1994PThPh..92..779T,
1995NuPhA.583..623B,Prak97,1994ApJ...436..257H,1997A&A...321..822G,
1998A&A...330.1005G,Stro99,1999A&AS..134...39S}.  
Because of the uncertainties of the composition of the
matter in the interior of neutron stars, the effects of some exotic
states, such as hyperonic matter \citep{Prak97,Pons99}, quark matter
\citep{Pons01prl}, kaon condensation \citep{Pons01} and the quark-hadron
phase transition \citep{Prak95}, on the evolution of PNSs have also
been explored.  Because the trapped neutrinos increase the chemical
potential of electrons, the common consequence of the inclusion of the
possible exotic states which contain negatively charged components in
the PNS is the existence of a metastable star, the mass of which is
larger than the maximum mass for a cold neutrino-free neutron
star. Inevitably, the metastable PNS will collapse to a black hole at
some time during the Kelvin-Helmholtz epoch.  Meanwhile, the neutrino
signal from the PNS will cease suddenly, because the time scale of the
collapse is on the order the free-fall time which is much shorter than
the evolutionary time scale of the PNS.  

As the trapped neutrinos escape, the global structure of the PNS
changes, especially the momentum of inertia which affects the stellar
rotational frequency dramatically.  Our key assumption is that 
Kelvin-Helmholtz timescale over which the neutrinos escape ($\sim
10$~s) is much longer than the dynamical timescale of the PNS
($\sim 1$~ms); therefore, the protoneutron star evolves through a
series of hydrostatic equilibria with successively smaller neutrino
fractions.  
In this work, we investigate the stellar global properties as 
a function of the number density of the trapped electron neutrinos.
Especially, 
we contrast the
rotational evolution of hyperonic PNSs with ordinary PNSs.  The
evolutionary behavior of a PNS in whose core the other possible
exotic states exist is qualitatively similar to the results for the
hyperonic star.  In order to describe the hyperonic matter with
trapped neutrinos, the relativistic mean field theory (RMFT) has been
generally applied \citep{Wale74,Chin77,Sero79,Sero86,1996NuPhA.606..508M,
Prak97,Yuan99,
Glen00}. The potential model is also used sometimes
\citep{Prak97}. Recently, the potential model with the
Bruecker-Bethe-Goldstone many-body theory at zero temperature was
employed to model the hyperonic matter with trapped neutrinos, and to
investigate the global properties of the PNSs. As expected, the
qualitative results which have been obtained in the both models are
very similar \citep{vida03}.  
The previous works have indicated that the effect of
the trapped neutrinos dominates over that of temperature on the
internal composition and EOS \citep{Prak97}. Thus, for simplicity, we
use the RMFT to describe the dense nuclear matter with hyperons at
zero temperature, and explore the response of the rotational frequency
to the escape of the neutrinos.

After its birth, the neutron star generally spins fast initially, even
near to the Keplerian angular velocity, $\Omega_{\rmscr{K}}$
\citep{2000ApJ...541.1033F}. Stellar rotation
is essential for the generation of the magnetic field 
\citep[\protect{\em e.g.}][]{1993ApJ...408..194T} and affects the
global properties of the PNS significantly \citep{Akiy03}.  In order
to to deal with the rotation of a rapidly rotating neutron star, a
fully general relativistic method should be used.  Several independent
techniques exist in the literature \citep{Ster98}.  Our calculation is
based on the Stergioulas \& Friedman's KEH codes 
\citep{1989MNRAS.237..355K,1989MNRAS.239..153K}, which is available
as a public domain code \citep{Ster95}.

This paper is organized as follows.  A brief description of the RMFT
for dense matter with hyperons, the microscopic properties of
dense matter in $\beta$-equilibrium, and the global properties of the
non-rotating and rapidly rotating PNSs are given in \S~\ref{sec:eos}.
In \S~\ref{sec:rota}, the numerical results of the evolution of the
rotational frequency is presented. The results and discussions will be
given in \S~\ref{sec:conl}.

\section{Equations of state for protoneutron stars}
\label{sec:eos}
\subsection{Description of the dense matter}

In the framework of relativistic field theory, the interaction
between nucleons and hyperons is mediated by the exchange of three
meson fields, $\sigma$, $\omega$, and $\rho$ mesons.  The coupling
constants and meson masses can be algebraically related to the bulk
properties of the symmetric nuclear matter \citep{Glen00}.
The Lagrangian of the hadrons is given by,
\def\vecrho{\mbox{\boldmath $\rho$}}
\def\vectau{\mbox{\boldmath $\tau$}}
\bea
\cal{L} 
  &=& \sum_{i}\overline{\psi}_{i} 
	\left [i\gamma_{\mu} \left (\partial^{\mu}
	+ig_{\omega i}\omega^{\mu}
	+i\frac{g_{\rho i}}{2} {\vectau} \cdot 
	{\vecrho^{\mu}} \right ) -(m_i-
	g_{\sigma i} \sigma)
 \right ] \psi_i \nonumber \\
  &+& \frac{1}{2} \left (\partial_{\mu} \sigma \partial^{\mu} \sigma
	-m_{\sigma}^2 \sigma^2  \right ) -U(\sigma) \nonumber \\
  &-& \frac{1}{4} \omega_{\mu \nu} \omega^{\mu \nu} 
	+\frac{1}{2}m_{\omega}^2 \omega_{\mu} \omega^{\mu} \nonumber \\
  &-& \frac{1}{4} \vecrho_{\mu \nu} \cdot \vecrho^{\mu \nu} 
	+\frac{1}{2}m_{\rho}^2 \vecrho_{\mu} \cdot \vecrho^{\mu},
\label{eq:L}
\eea
where,
\bea
\omega_{\mu\nu}&=&\partial_{\mu}\omega_{\nu}-\partial_{\nu}\omega_{\mu}, \\
\vecrho_{\mu\nu}&=&\partial_{\mu}\vecrho_{\nu}-\partial_{\nu}{\vecrho}_{\mu}
  +g_{\rho n}\vecrho_{\mu} \times \vecrho_{\nu}.
\eea
Throughout this paper we will use the set of natural units where
$\hbar=c=1$ unless noted otherwise. 
Here $\psi_i$, $\sigma$, $\omega$, and $\vecrho$ denote the fields of the
baryon of species $i$ ($i=n,p,\Lambda, \Sigma^-, \Sigma^0, \Sigma^+,
\Xi^-, \Xi^0$), the mesons $\sigma$, $\omega$, and $\rho$ with the
masses of $m_i$, $m_{\sigma}$, $m_{\omega}$, $m_{\rho}$,
respectively. The constants $g_{\sigma i}$, $g_{\omega i}$, $g_{\rho
i}$ are coupling constants for interactions between mesons and
baryons.  $\vectau=(\tau_1,\tau_2,\tau_3)$ denotes the $2\times 2$ Pauli 
isospin matrices and the dot and cross products are calculated over isospin space.
The potential of the self-interaction of the scalar field
which reads,
\be
U(\sigma)=\frac{1}{3}bm_n(g_{\sigma n}\sigma)^3 
	+ \frac{1}{4}c(g_{\sigma n}\sigma)^4,
\label{eq:U}
\ee
is introduced to produce reasonable incompressibility of nuclear
matter \citep{Bogu77}. Here the coefficients $b$ and $c$ denote
self-coupling constants for the $\sigma$ meson field.

The dynamical equations of nucleons and mesons can be derived from the
above Lagrangian. Generally, these equations are coupled.  Applying
the RMFT approximation \citep{Wale74}, the dynamical equations are
decoupled, they are,
\bea
\mu_i    &=&g_{\omega i} \omega_0 +g_{\rho i} \rho_{03} I_{3i}
	+\sqrt{k^2+m^{*2}_i}, \\
m_{\sigma}^2 \sigma &=& g_{\sigma i} n_{{\rm s}i} 
	-\frac{\partial U(\sigma)}{\partial \sigma}, \\
\omega_0 &=& \sum_{i}\frac{g_{\omega i}}{m^2_{\omega}} n_i, \\
\rho_{03} &=& \sum_{i}\frac{g_{\rho i}}{m^2_{\rho}} I_{3i} n_i.
\label{eq:rho}
\eea
Here, $\mu_i$ is the chemical potentials of baryons,
$I_{3i}$ is the third component of isospin for the baryons, and
$m^*_i=m_i-g_{\sigma i} \sigma$ are the effective masses of baryons. 
Finally, the scalar density $n_{{\rm s} i}\equiv \left <\overline{\psi} \psi \right >$ 
is defined as
\be
n_{{\rm s}i}=\frac{2}{\pi^2} \int_0^{k_i^{\rm F}} \frac{m^*_i}
	{\sqrt{k^2+(m^*_i)^2}} k^2 dk,
\ee
where $k_i^{\rm F}$ are the Fermi momenta of baryons.
The condition of chemical equilibrium between baryons and leptons
is given by,
\be
\mu_{i}=b_i\mu_n-q_i\tilde{\mu_l},
\ee
where $b_i$ is the baryon number of particle $i$ and $q_i$ is 
its charge. For the neutrino-free case, $\tilde{\mu_l}=\mu_l$, 
for the trapped-neutrino case, $\tilde{\mu_l}=\mu_l-\mu_{\nu_l}$.
Here $\mu_l$ $(l=e^-, \mu^-)$ are 
the chemical potentials of the leptons.
Because neutrinos are trapped, the lepton number per baryon
$Y_{{\rm L}_l}$ of each flavor must be conserved
on dynamical time scales,
\be
Y_{{\rm L}_l}=Y_l+Y_{\nu_l}=\hbox{\rm constant}.
\ee
Studies of the gravitational collapse calculations of the core of a
massive star indicate that $Y_{{\rm L}_e}\simeq 0.4$ 
\citep{Prak97}. In addition,
because no muons appear when neutrinos become trapped, $Y_{{\rm
L}_{\mu}}=0$.  During the Kelvin-Helmholtz epoch, $Y_{\nu_e}$ changes
from its initial value to zero.  Therefore, if the chemical potential
of neutrons and electrons are known, the fields of the mesons $\sigma$,
$\omega$, and $\rho$ can be solved numerically, and then the momenta
of all the baryons can be determined at the same time.  Given the
baryon number density $n_{\rm B}$, the additional condition, charge
neutrality, is needed to determined $\mu_n$ and $\mu_e$
self-consistently.

According to the standard procedure of RMFT theory \citep{Wale74}, 
the energy density and the pressure of the baryons can be obtained,
\bea
\epsilon_{\rm B}&=&\frac{1}{2}m_{\sigma}^2\sigma^2+\frac{1}{2}m_{\omega}^2
\omega_0^2
+\frac{1}{2}m_{\rho}^2\rho_{03}^2
+\sum_{i}\frac{1}{\pi^2} \int_{0}^{p^{\rm F}_i}\sqrt{p^2+m_i^*}p^2dp,
\label{eq:e} \\ 
p_{\rm B}&=&-\frac{1}{2}m_{\sigma}^2\sigma^2+\frac{1}{2}m_{\omega}^2
\omega_0^2
+\frac{1}{2}m_{\rho}^2\rho_{03}^2
+\sum_{i}\frac{1}{\pi^2} \int_{0}^{p^{\rm F}_i}\frac{p^4dp}{\sqrt{p^2+m_i^*}}.
\label{eq:p}
\eea

The coupling constants $g_{\sigma N}$, $g_{\omega N}$, $g_{\rho N}$,
$b$, $c$ ($N=n,p$) in RMFT are algebraically related to the bulk properties of
symmetric nuclear matter at saturation density. In our calculation,
we choose the so-called GM1 set \citep{Glen91},
\be
\left (\frac{g_{\sigma N}}{m_{\sigma}}\right )^2=11.79, ~~
\left (\frac{g_{\omega N}}{m_{\omega}}\right )^2=7.149, ~~
\left (\frac{g_{\rho N}}{m_{\rho}} \right )^2=4.411, ~~
b=2.947 \times 10^{-3},~~
c=-1.07 \times 10^{-3}~~,
\ee
from which the nuclear properties arise as follows: the
saturation density, effective mass of nuclear, incompressibility, and
binding energy per nucleon are $n_0=0.153$~fm$^{-3}$, $m^*=0.7$,
$K=300$~MeV, and $B/A=-16.3$~MeV.

At the saturation density, the bulk properties are independent of the
hyperon couplings which could be determined by reproducing the binding
of the $\Lambda$ hyperon in nuclear matter \citep{Glen91}.  It is
assumed that all hyperons in the octet have the same coupling which
are expressed in terms of nucleon couplings,
\be
x_{\sigma H}=\frac{g_{\sigma H}}{g_{\sigma N}},~~
x_{\omega H}=\frac{g_{\omega H}}{g_{\omega N}},~~
x_{\rho H}=\frac{g_{\rho H}}{g_{\rho N}}.
\ee
In the GM1 set \citep{1997PhRvL..79.1603G,Glen91,Glen00}, 
\be
x_{\sigma H}=0.6, ~~
x_{\omega H}=x_{\rho H}=0.653.~~
\ee

\subsection{The interior composition and the EOS}

The evolution of the PNS with time should be investigated by solving
the neutrino transport equations \citep{Pons99,2003fthp.conf...39J,
2002A&A...396..361R}.  Due to the escape of
the neutrinos, $Y_{\nu_e}$ decreases with time. Because our purpose in
this work is to explore the rotational behavior of the PNS as the
neutrinos escape, we study the rotational frequency and other
interesting quantities as a function of $Y_{\nu_e}$. The ratio of the
number fraction of electron neutrinos to its initial value is assumed
to be a constant with baryon number density, for a given time. Because
the density in the stellar core is almost constant, this is a good
approximation.  As the electron neutrinos escape, the muon neutrino
appears due to the following reaction,
\be
e^- \rightarrow \mu^-+\overline{\nu}_{\mu}+\nu_e ~~.
\ee 
To describe the escape of the muon neutrinos, we assume that the ratio
of the number of the escaped muon neutrinos to that of the muon neutrinos which
should escape freely if the matter is transparent to neutrinos equals
to $Y_{\nu_e}/Y_{\nu_e}^i$. Our
calculations show the number density of muons is generally several
orders of magnitude lower than that of electrons, so its effects on
the EOS of PNS could be neglected, even though we have taken it into
consideration.

\begin{figure}
\epsscale{0.8}
\plotone{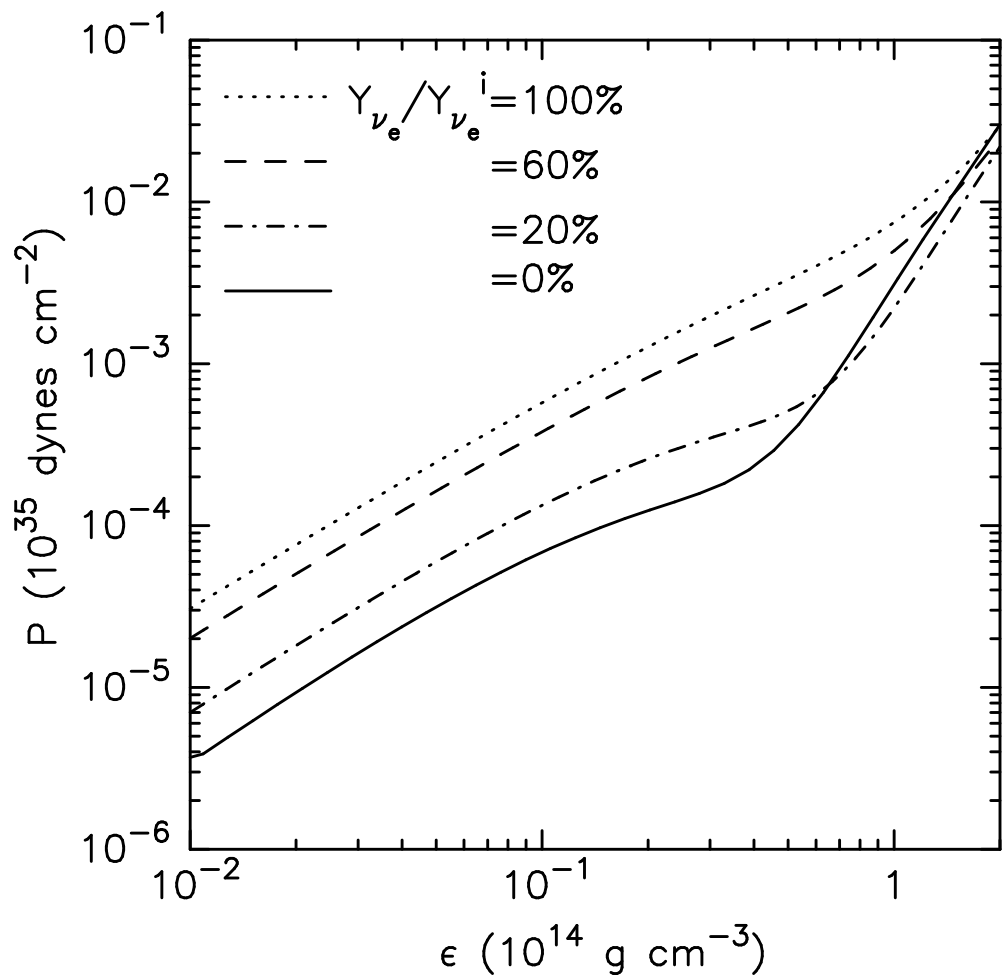}
\caption{The pressure versus the energy density in the stellar envelope
at different ratios of the fraction of trapped electron neutrinos
to the initial value. Before the neutrinos escape, $Y_{{\rm L}_e}=0.4$,
$Y_{{\rm L}_{\mu}}=0$.}
\label{fig:crust}
\end{figure}

Previous studies have shown that neutrinos are trapped above a density
$n_{\rm env}\approx 6\times 10^{-4}$~fm$^{-3}$
\citep{1995ApJ...450..830B,Stro99}.  Therefore, before the neutrinos
escape, we set $Y_{\rm L_e}=0.4$ for densities above $n_{\rm env}$.
For the description of neutrino-free subnuclear matter, the FPS EOS
\citep{1993PhRvL..70..379L} is chosen, which is smoothly matched to
the EOS of nuclear matter described in RMFT.  At low baryon number
density ($n_{\rmscr{B}}<0.1$ fm$^{-3}$), the nucleons are
non-relativistic, thus the pressure contributed by the trapped
neutrinos and the relativistic eletrons dominates.  The EOS of
subnuclear dense matter becomes much stiffer due to the trapped neutrinos
(see Fig.~\ref{fig:crust}).

\begin{figure}
\epsscale{0.8}
\plotone{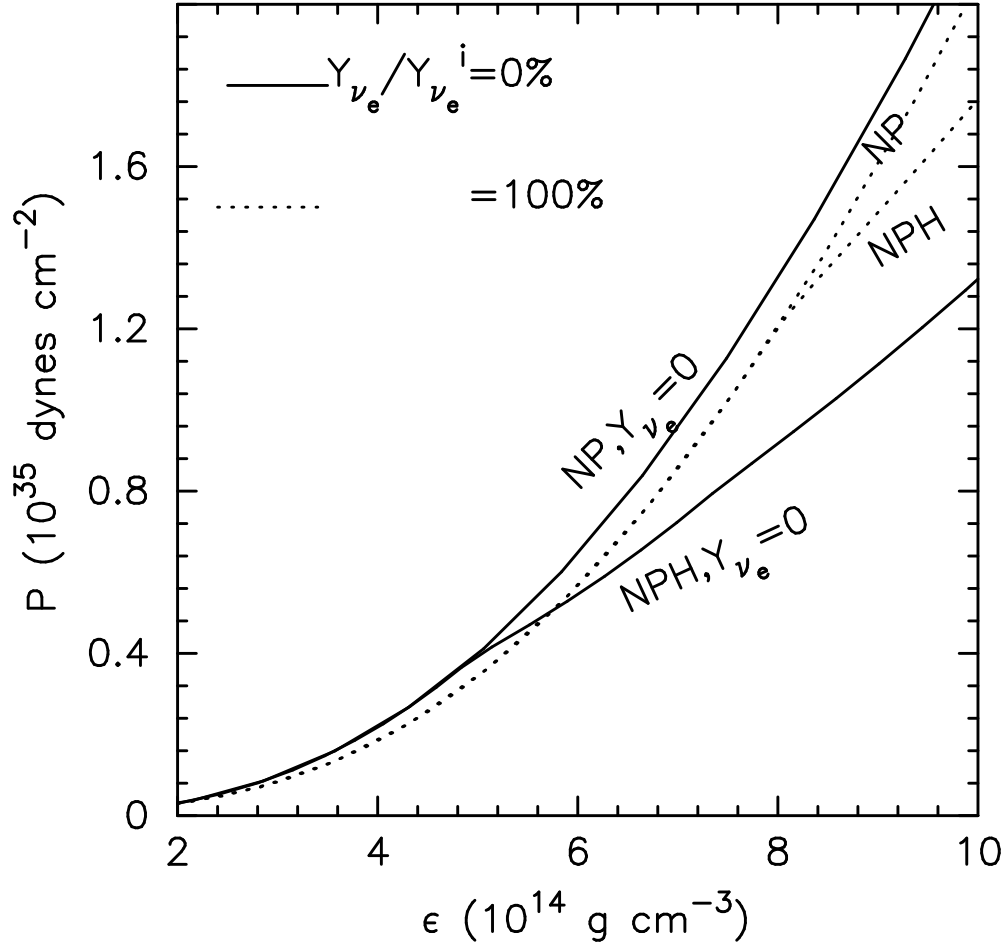}
\caption{The comparison of the EOSs between the neutrino-trapped matter
(dotted lines) and the neutrino-free matter (solid lines) in
a stellar core which contains only nucleons (labeled as 'NP')
and which contains hyperons as well (labeled as 'NPH').}
\label{fig:eos}
\end{figure}

The main effect of the neutrino trapping in the dense matter above 
nuclear density is that the fraction of electrons increases
dramatically, regardless of the interior composition.  Therefore, for
a protoneutron star which contains only nucleons in its core, the
fraction of protons is forced to increase in the same manner because
of charge neutrality.  Comparing to the neutrino-free case, the
symmetry energy which is contributed by the $\rho$ meson and linearly
depends on the difference between the baryon number density of
neutrons and protons, decreases significantly (see Eq.~\ref{eq:rho}
and Eq.~\ref{eq:e}-\ref{eq:p}).  Consequently, the EOS in the
normal stellar core is softer with trapped neutrinos than without
them.  For hyperonic matter, the higher chemical potential of
electrons that results from trapped neutrinos increases the critical
baryon number density of the appearance of $\Sigma^-$ particles.
Because there are fewer species of baryons, the Fermi momenta of each
species is larger, so the pressure of hyperonic matter with trapped
neutrinos is larger than the pressure without neutrinos (see
Fig.~\ref{fig:eos}).

\begin{figure}
\epsscale{0.8}
\plotone{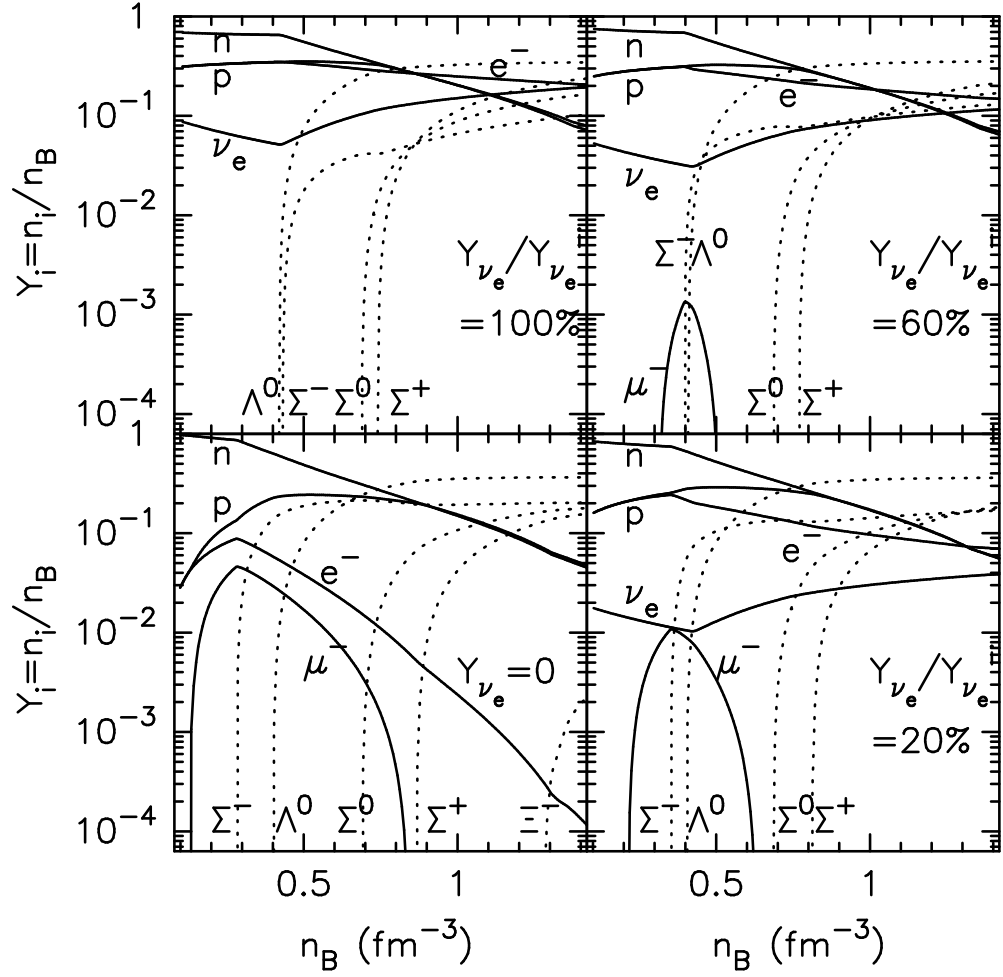}
\caption{The relative compositions ($Y_i$) as a function of the baryon
number density ($n_{\rm B}$) at the different values of $Y_{\nu_e}/Y_{\nu_e}^i$
in the hyperonic matter.}
\label{fig:comp_npH}
\end{figure}

\begin{figure}
\epsscale{0.8}
\plotone{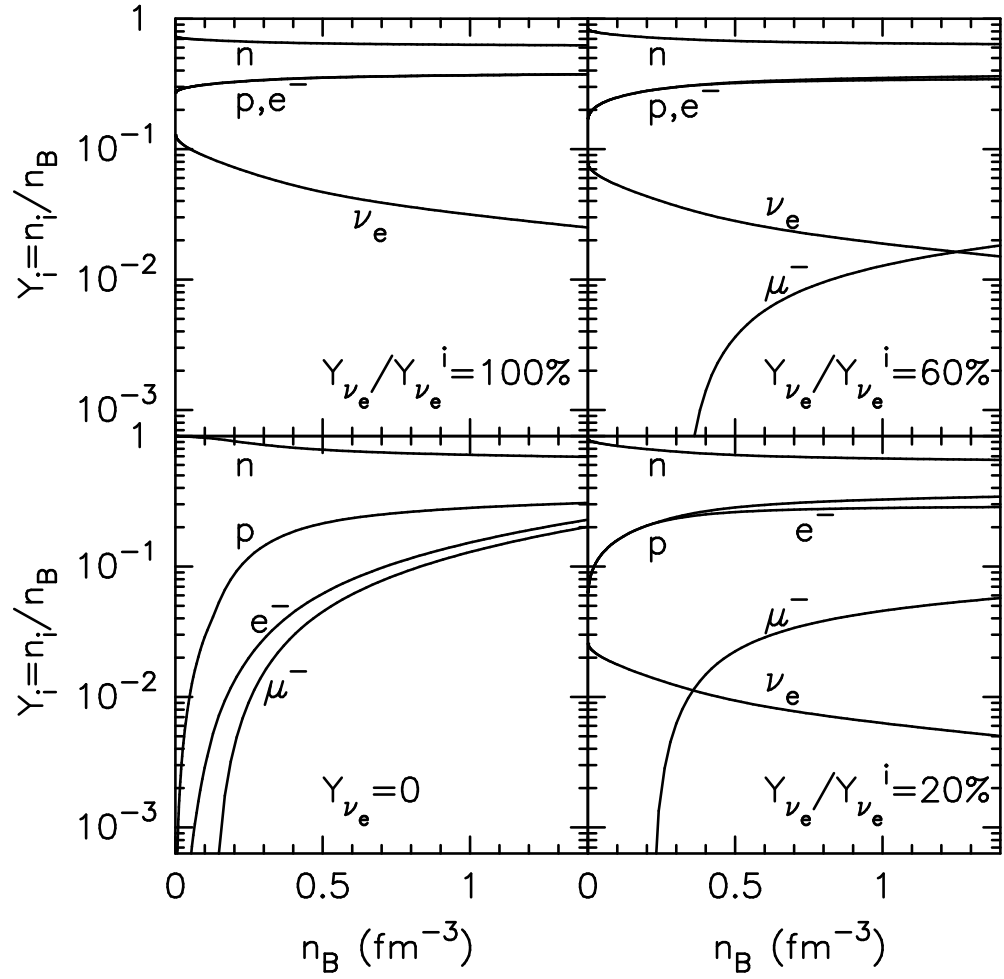}
\caption{The same as Fig.~\ref{fig:comp_npH}, but for the normal nuclear matter
which consists of nucleons. }
\label{fig:comp_np}
\end{figure}

Figure~\ref{fig:comp_npH} shows the composition of matter in the
stellar interior which contains hyperons.  The subfigures in
Fig.~\ref{fig:comp_npH} are the results for $Y_{\nu_e}/Y_{\nu_e}^i=100\%,
60\%, 20\%, 0$, respectively. As noticed in previous work, the
neutrino trapping causes the following results: (1) muons do not
appear; (2) the number density of electrons and protons increases
significantly; (3) the onset of the negatively charged hyperons takes
place at higher baryon number density, chargeless hyperons
appear at nearly the same baryon number density, while the onset of positively
charged hyperons occurs at lower density (see Fig.~\ref{fig:comp_npH}). 
The whole effect is that the EOS of the hyperonic matter 
becomes stiffer at high density.
With the escape of the trapped neutrinos, muons begin to appear and
$\Sigma^-$ particles appear at lower density.

For the purpose of comparison, Fig.~\ref{fig:comp_np} shows the
composition of the matter which includes only nucleons. As in the
above case, the neutrino trapping suppresses the appearance of muons
and increases the number density of electrons and protons.

\begin{figure}
\epsscale{0.8}
\plotone{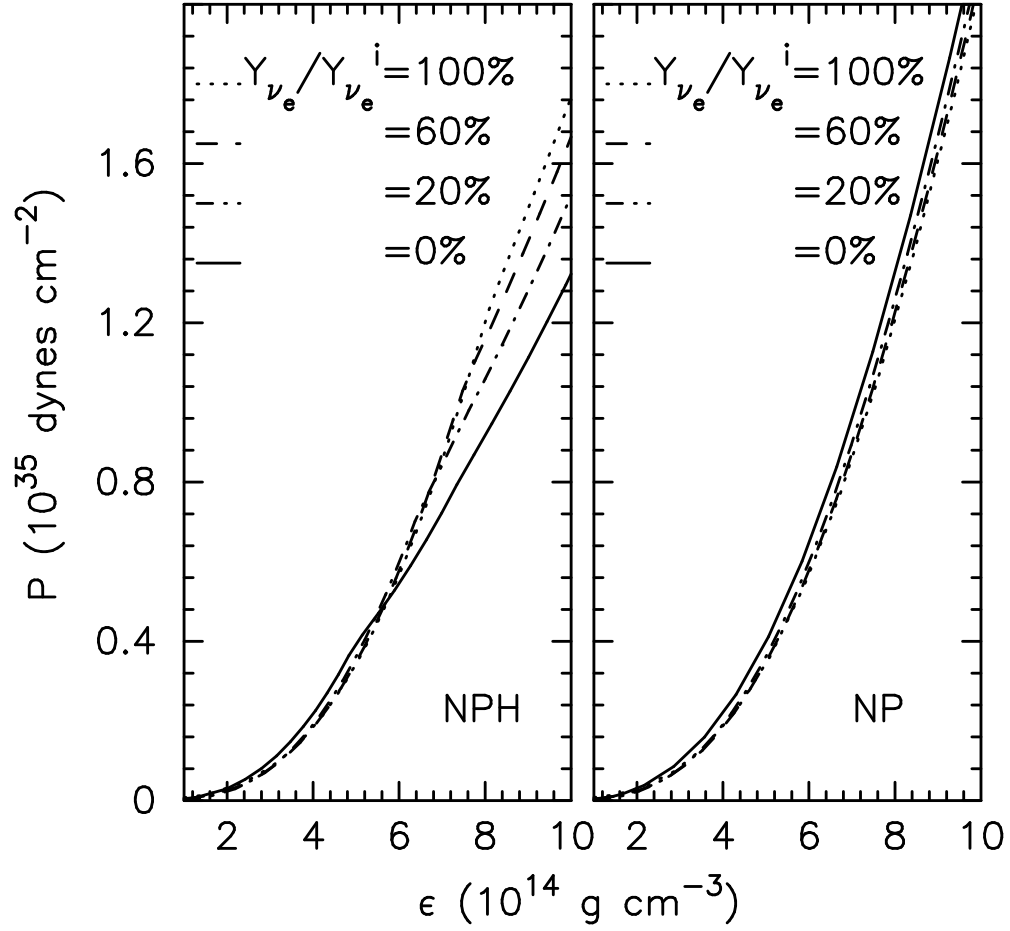}
\caption{The change of the EOSs of the hyperonic matter (left panel),
and the normal nuclear matter (right panel) due to 
the escape of the neutrinos.  }
\label{fig:eos_npH}
\end{figure}

\begin{figure}
\epsscale{0.8}
\plotone{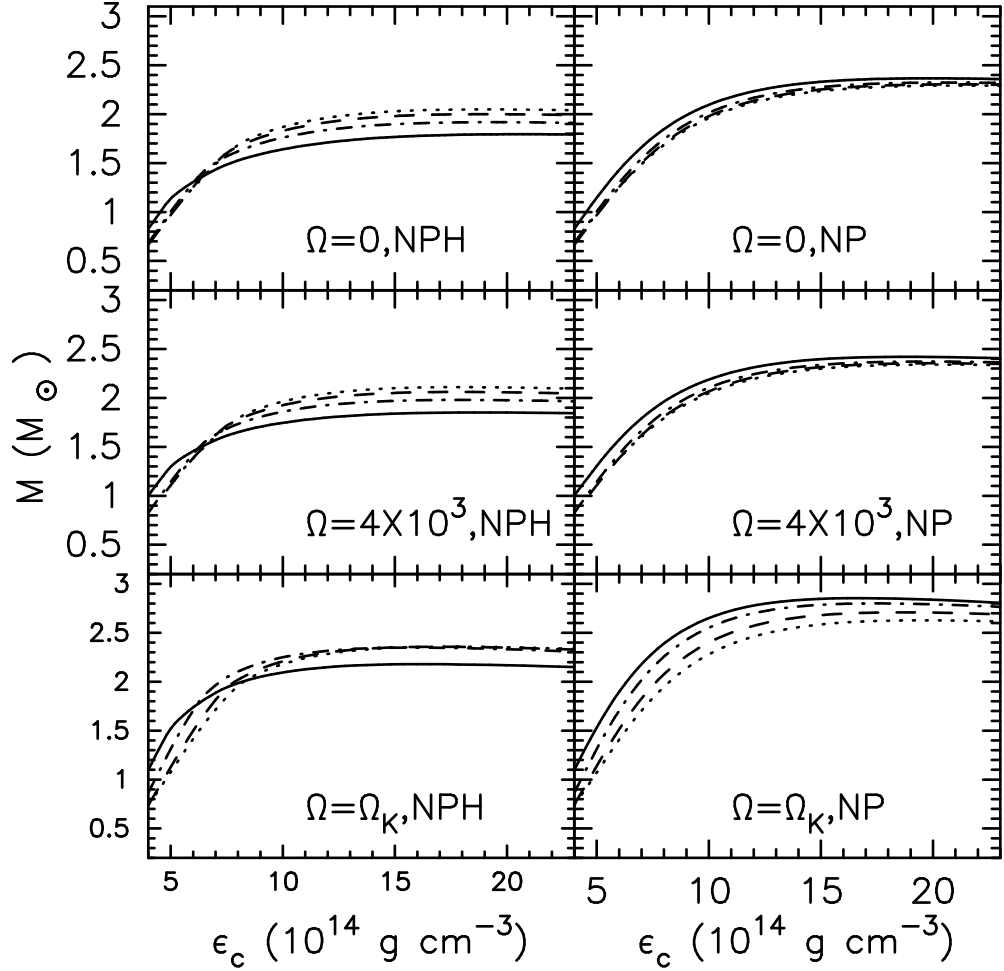}
\caption{The gravitational mass $M$ as a function of the central
density at the different rotation frequencies $\Omega=0$,
$\Omega=4 \times 10^3$, $\Omega=\Omega_{\rm K}$.
Left panels are the results for hyperonic matter, while
right panels for the nuclear matter. The solutions traced
by the various lines are the same as in Fig.~\ref{fig:eos_npH}, which 
represent the
different values of $Y_{\nu_e}/Y_{\nu_e}^i$. }
\label{fig:mr_m}
\end{figure}

\begin{figure}
\epsscale{0.8}
\plotone{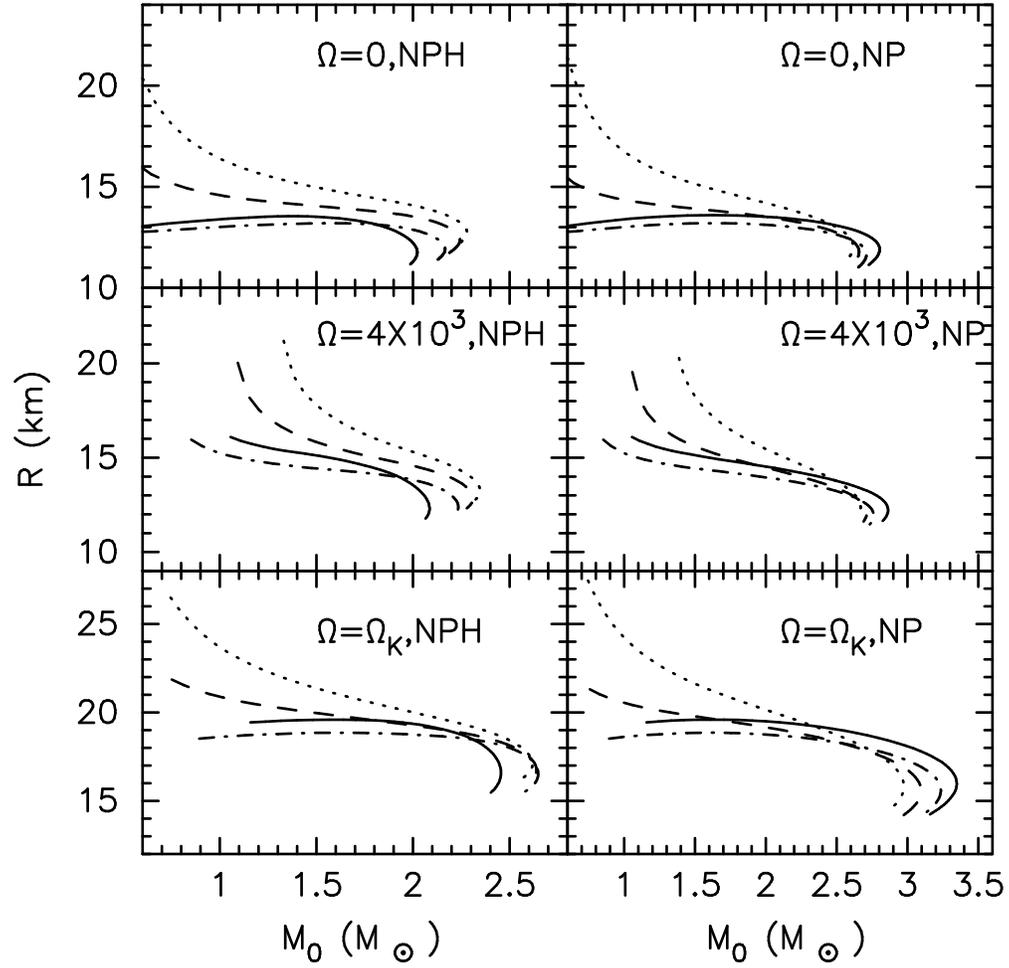}
\caption{The stellar radii as a function of the baryonic mass.
The solutions traced  by the various lines are the same as in Fig.~\ref{fig:eos_npH}.
}
\label{fig:mr_r}
\end{figure}

\begin{figure}
\epsscale{0.8}
\plotone{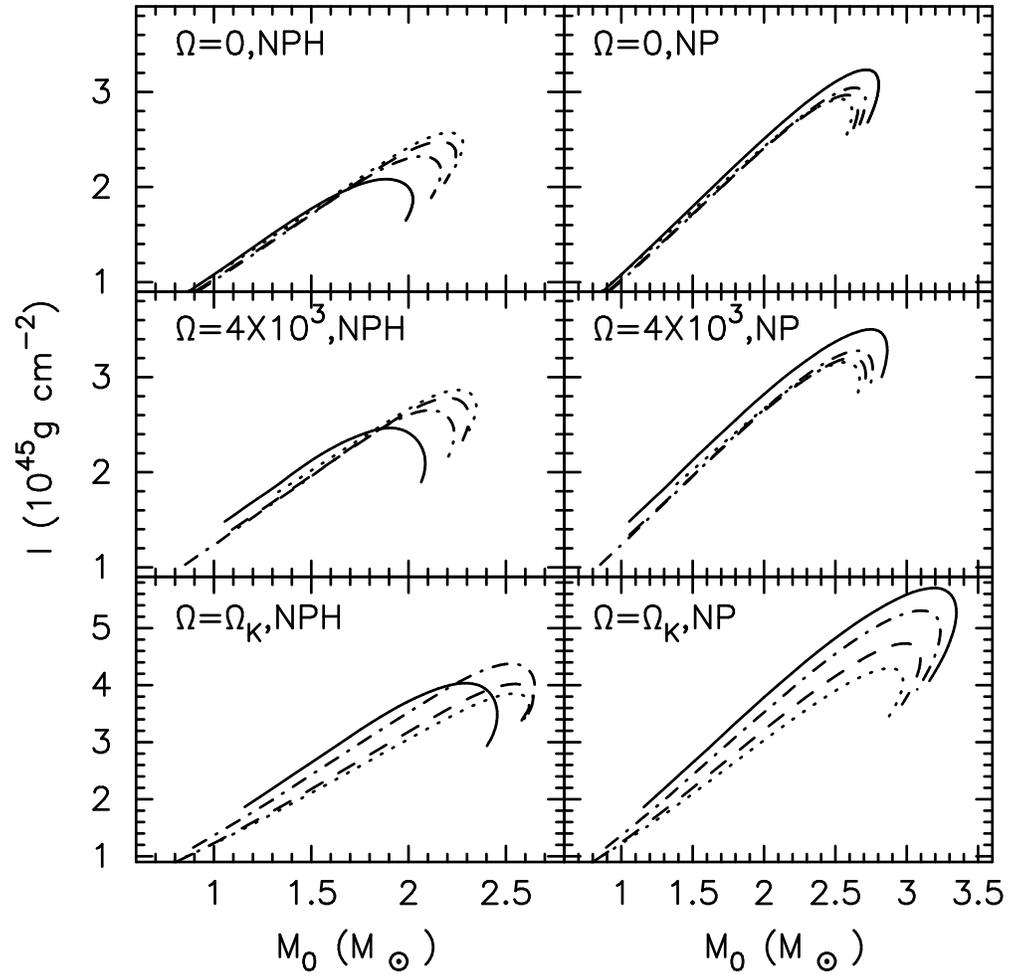}
\caption{The stellar momenta of inertia as a function of the baryonic mass.
The solutions traced  by the various lines are the same as in Fig.~\ref{fig:eos_npH}.}
\label{fig:mr_i}
\end{figure}

The responses of the EOSs based on the hyperon model (NPH) and nucleon
model (NP) as the neutrinos escape are shown in
Fig.~\ref{fig:eos_npH}. In the NP model, as expected, the EOSs become
more and more stiff as the neutrinos escape.  The same occurs in the
NPH model, below the critical density of the emergence of the
hyperons.  However, in the NPH model, at densities above the critical
density, the EOSs become softer due to the appearance of the more and
more additional components.

\subsection{Properties of the non-rotating and rapidly rotating PNSs}

In order to solve the Einstein field equations with the source terms,
neutron-star matter is generally assumed to behave as a perfect fluid. 
For a static neutron star, the resulting equations are well known, the
Tolman-Oppenheimer-Volkoff equations. It is complicated to deal with
the rotation of a neutron star. If the stellar angular velocity
$\Omega$ is small compared to the critical value $\Omega_0=\sqrt{4\pi
G\epsilon_c}$, where $\epsilon_c$ is the mass density at the center of the
star, the perturbation theory developed by Hartle and Thorne is
accurate enough \citep{Hart67,Hart68,Chub00}. Hartle's original 
perturbation theory was improved by considering the effects of
centrifugal stretching and frame dragging in an attempt to calculate the
properties of a rapidly rotating neutron star with $\Omega \sim
\Omega_0$ \citep{Glen92a,Glen92b}.  Nevertheless, the exact treatment
of the rotation of a compact star should be done in full general
relativity.  Several reasonable assumptions are generally made, they
are, the spacetime and the matter are stationary and axisymmetric, and the
star uniformly rotates. Under these assumptions, the metric of the
spacetime can be expressed by 
\be
ds^2=e^{\gamma+\rho}dt^2+e^{2\alpha}(dr^2+r^2d\theta^2)
+e^{\gamma-\rho}r^2\sin^2\theta(d\phi-\omega dt)^2,
\label{eq:axisym}
\ee 
where the metric potentials $\gamma, \rho, \omega$ and $\alpha$ are
functions of $r$ and $\theta$ only.  Several independent methods to
numerically calculate the metric potentials exist.  After solving the
metric potentials, any coordinate-invariant physical quantity can be
calculated.

Figures~\ref{fig:mr_m}-\ref{fig:mr_i} show the global properties of
the neutron stars.  In Fig.~\ref{fig:mr_m}, we plot the gravitational
mass ($M$) as a function of the central energy-density ($\epsilon_c$)
for PNSs with different fractions of trapped neutrinos
$Y_{\nu_e}/Y_{\nu_e}^i$ and with different rotational frequencies
$\Omega=0$, $\Omega=4 \times 10^3$, and $\Omega=\Omega_{\rm K}$ in
both the NP and NPH model.  As predicted in the previous works, the
maximum mass of the static PNS decreases as the neutrinos escape in
the NPH model, which results in the possible existence of a metastable
star.  In the NP model the maximum mass increases as the neutrinos
escape.  These results hold even when the PNS spins very rapidly.

Figures~\ref{fig:mr_r}-\ref{fig:mr_i} show the radii and the moments
of inertia as a function of the rest mass respectively.  In the NP
model, if the stellar rest mass $M_0$ is near the maximum mass, the
radii $R$ and the moments of inertia $I$ increase as the neutrinos
escape, while decrease in the NPH model. Evidently, this behavior is
dominated by the change of the EOSs in the stellar core.  The trapped
neutrinos always make the EOS in the crust stiffer (see
Fig.~\ref{fig:crust}) because of the dominance of the lepton
pressure.  When the initial mass is small, firstly, the PNS shrinks
as the neutrino escape the crust, then begins to expand when
the pressure of the $\rho$ meson which determines the symmetry energy
is comparable to that of the leptons.

Figure~\ref{fig:mr_i} show the moments of inertia as a function of the
rest mass. In the NP model, if the stellar rest mass $M_0$ is near the
maximum mass, the moments of inertia $I$ increase as the neutrinos
escape, as the radii $R$ do.  In the NPH model, the moments of
inertia decrease.  When the initial mass is small, firstly, $I$
decreases then increases while the neutrinos escape.  The change in
behavior is not as significant as what happens to the radii, because
gravitational mass increases as the neutrinos escape the core.  In
some sense, this compensates the decrease of the radius.

\section{Rotational evolution of protoneutron stars}
\label{sec:rota}
As investigated in the above section, the global structure of the PNS
changes in different ways as the neutrinos escape depending on whether
hyperons exist in the core.  This may affect the rotational evolution
of the PNS.  During the evolution of the PNS, the baryonic mass is
fixed, if there is no accretion. Neglecting the loss of angular
momentum due to the radiation of gravitational and electromagnetic
waves, the loss of the total angular momentum of the PNS only results
from the escaping of the neutrinos.  To determine how the loss of
angular momentum carried away by the neutrinos affects the rotational
evolution, we also assume that the neutrinos do not carry any angular
momentum as they escape.

\begin{figure}
\epsscale{0.8}
\plotone{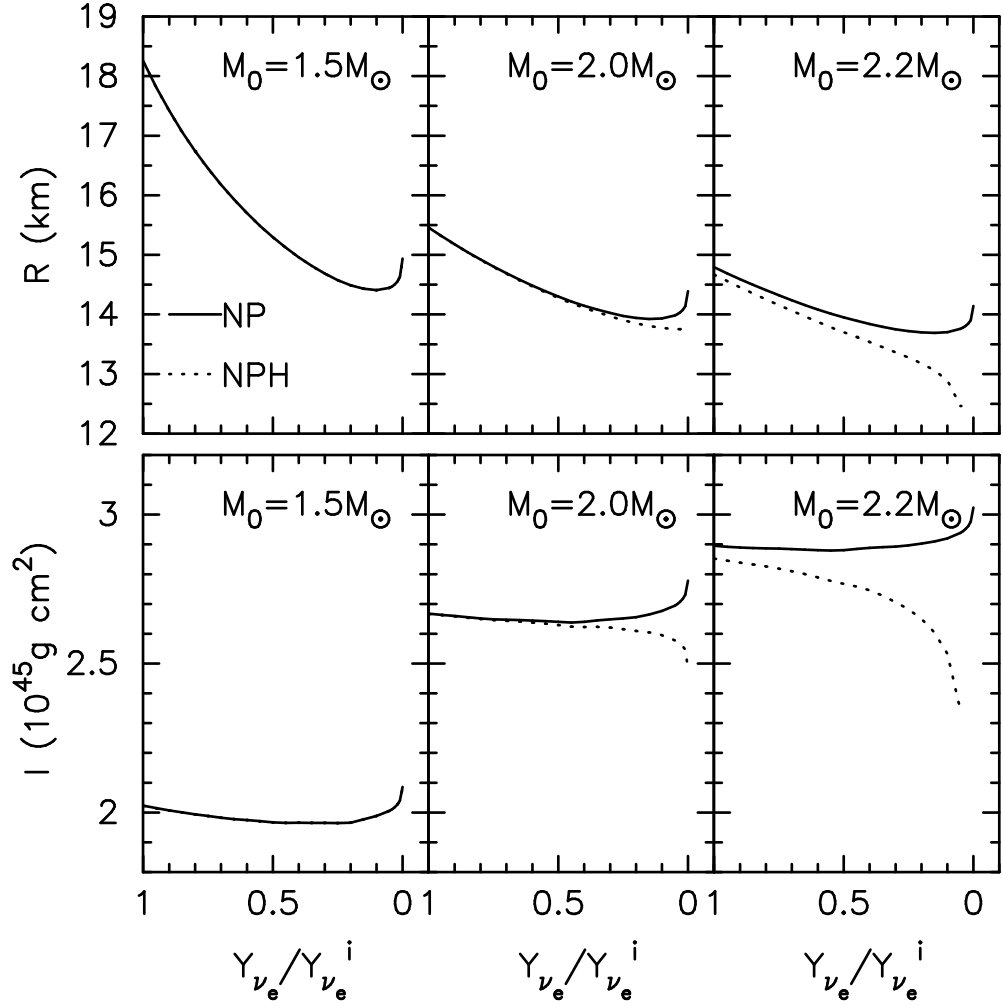}
\caption{The evolution of the stellar radii (upper panels) and the
momenta of inertia (lower panels)
with the escape of the neutrinos at the different fixed
rest masses $M_0=1.5,2.0,2.2M_{\sun}$. Dotted lines are
the results for the PNS with hyperons and solid lines
the PNS without hyperons. 
The initial spin is taken to be $\Omega=4000$~s$^{-1}$.}
\label{fig:evol_ri}
\end{figure}

Figure~\ref{fig:evol_ri} shows the stellar radii and moment of inertia
as a function of neutrino fraction.  For the NP model, the radii
decrease first mainly because the EOSs of the subdense matter in the
stellar crust becomes softer as the neutrinos escape.  The radii then
increase mainly because the EOS of the dense matter in the stellar
core becomes stiffer, as contribution from symmetry energy increases.
For the NPH model, the radii keep decreasing because both the decrease
of the neutrino pressure and the appearance of more and more hyperons
in the stellar core make the EOS softer.  If $M_0=1.5M_{\sun}$,
hyperonic matter does not appear in the stellar core at all. If
$M_0=2.2M_{\sun}$, the PNS is metastable, and it collapses to a black
hole at the end of the Kelvin-Helmholtz epoch.  The evolution of the
total moment of inertia is very similar to that of the radii.

\begin{figure}
\epsscale{0.8}
\plotone{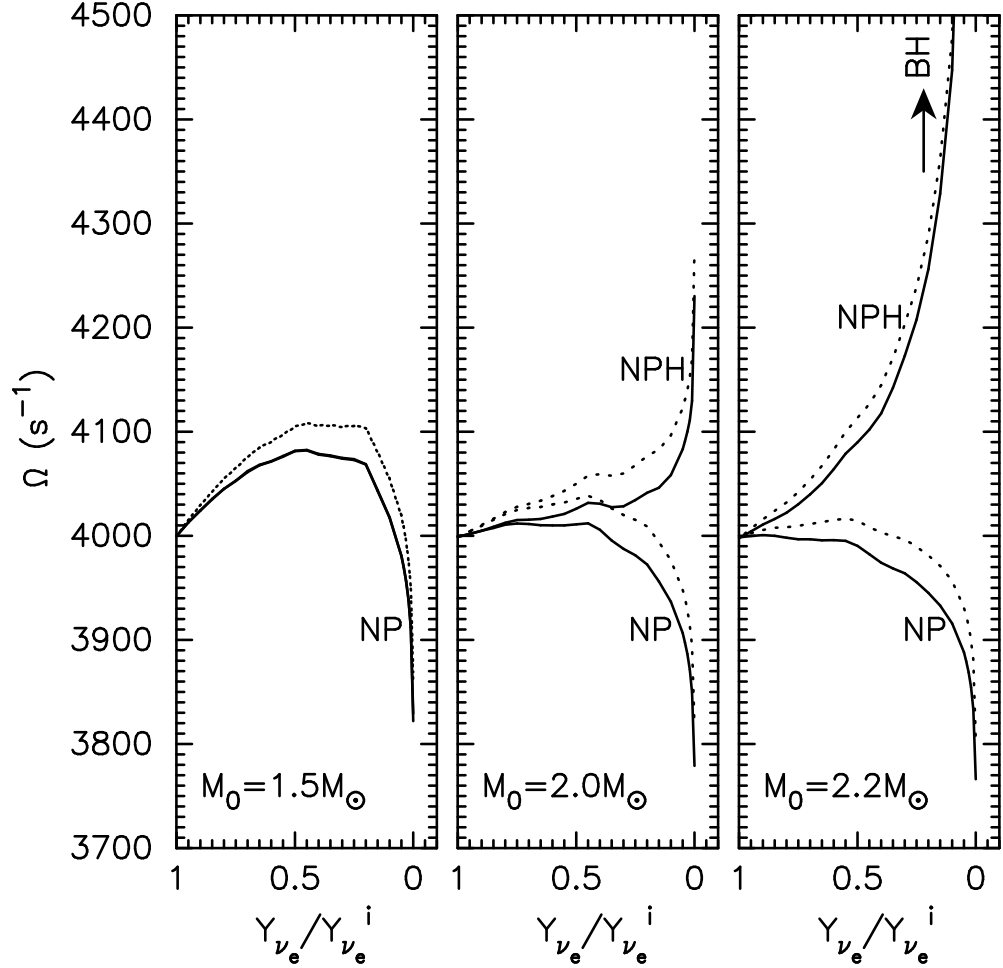}
\caption{The evolution of the rotation frequency of a PNS with hyperons
or not at the different fixed rest masses of the star,
$M_0=1.5,2.0,2.2M_{\sun}$.
The dotted lines represent the corresponding results for the
cases in which the stellar angular momentum carried away
by the escaped neutrinos is ignored. }
\label{fig:evol_o}
\end{figure}

Figure~\ref{fig:evol_o} shows how the rotational frequency changes as
the neutrinos escape in the NPH and NP model.  Early in the
Kelvin-Helmholtz epoch, the PNS always spins up independent of the
presence of hyperons, because as neutrinos escape the envelope, the
PNS shrinks. 
The crustal EOS affects
the rotational evolution of a protoneutron star significantly 
\citep{chen02}. 
Later in the evolution, roughly speaking, for the NPH
model, the PNS spins up with the EOS in the stellar core becoming
softer; while for the NP model, the PNS spins down with the EOS in the
stellar core becoming stiffer.  For $M=2.0M_{\sun}$, in the middle of
the evolution, the NPH spins down for a short period.  
Even though the increased number of hyperons make the EOS softer,
at the same time the fraction of the proton decreases, which makes the
EOS stiffer. These two effects compete with each other, the spin down
results from the increase of the symmetry energy. If the PNS is
metastable, the star keeps spinning up significantly, and then
collapses to a black hole at the end of its life.  It is clearly shown
in Fig.~\ref{fig:evol_o} that even though the escaped neutrinos carry
away some of the angular momentum which decreases the rotational
frequency of the PNS, the effect of the change of the global structure
due to the escape of the neutrinos dominates the stellar rotational
evolution.

\begin{figure}
\epsscale{0.8}
\plotone{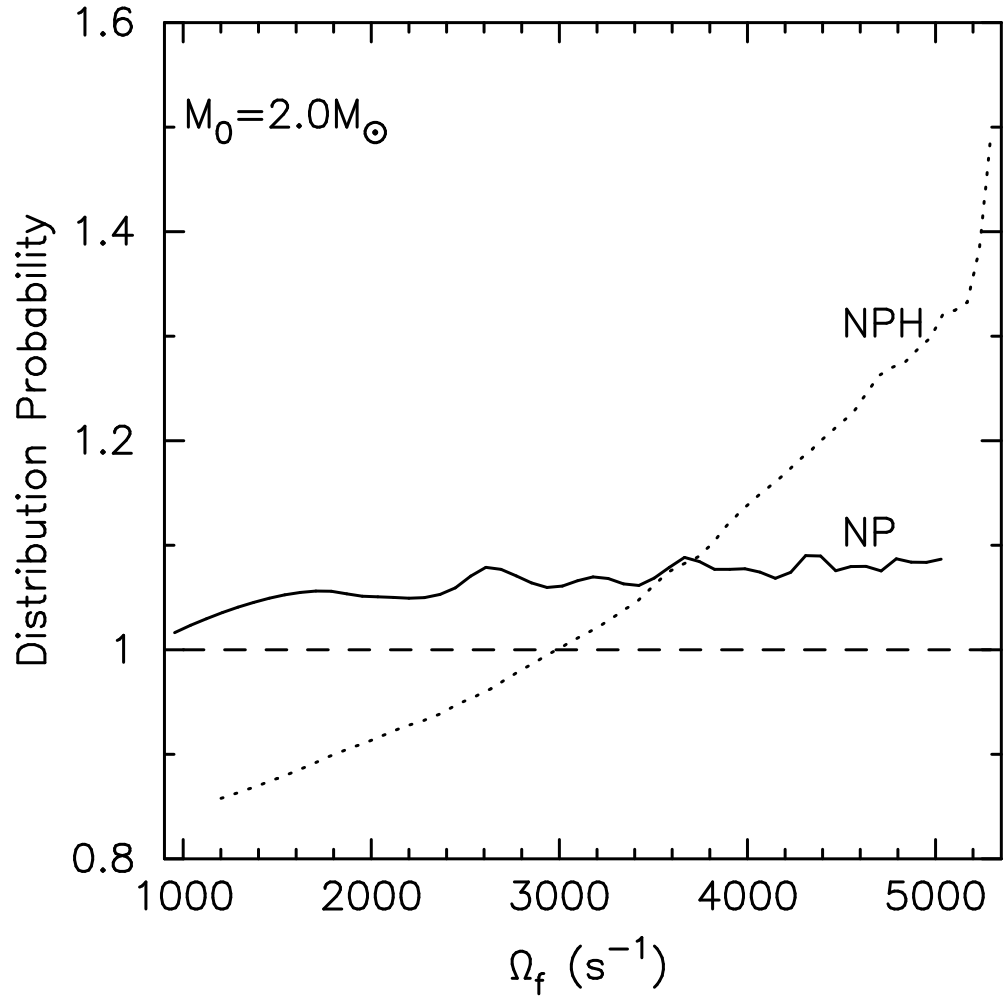}
\caption{The distribution probability (arbitrary units)
of the rotation frequencies of the PNSs
whose rest mass is 2.0$M_{\sun}$ at the end of the
Kelvin-Helmholtz epoch. The dashed line represents the assumed
uniform distribution of the initial spin periods. }
\label{fig:dist}
\end{figure}

The change of the rotational frequencies of the PNS after the neutrinos
escape would change the initial distribution of the periods of neutron
stars.  Assuming the distribution at the beginnning of the
Kelvin-Helmholtz epoch is uniform between zero and the Keplerian
frequency ($\sim 5344 s^{-1}$ for $M_0=2.0 M_{\sun}$), the new
distribution is shown in Fig.~\ref{fig:dist}.  For PNSs without
hyperons, the range of the periods is somewhat narrower than before, but
roughly speaking, the distribution is still almost uniform.  For the
PNS with hyperons, a significant number of slowly rotating PNSs shift
to higher spins; consequently, the presence of hyperonic matter at the
center of neutron stars will skew the distribution of initial spin
periods toward shorter periods.

\begin{figure}
\epsscale{0.8}
\plotone{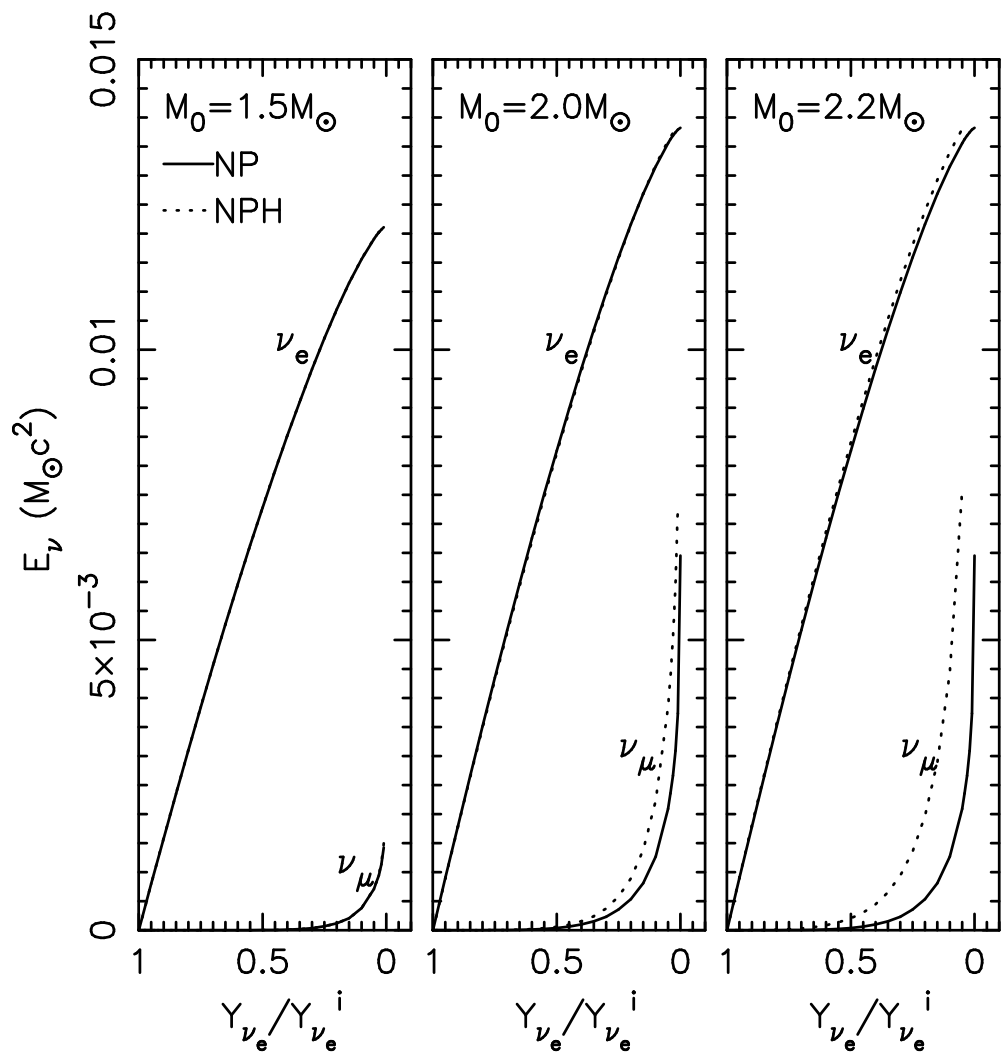}
\caption{The total neutrino energy radiating from the PNS as a function
of $Y_{\nu_e}/Y_{\nu_e}^i$. The symbol of the lines are the same
as Fig.~\ref{fig:evol_o}. }
\label{fig:neut}
\end{figure}

Figure~\ref{fig:neut} shows the evolution of the integrated flux of the 
escaped electron neutrinos. The total energy of the trapped 
electron neutrinos $E_{\nu}^{\rm tot}$ may be estimated as follows: 
the average baryon number density $n_{\rm B}\simeq M/(\case{4}{3}\pi R^3m_n)$, 
the corresponding neutrino number density $n_{\nu_e} =n_BY_{\nu_e}$, and the
Fermi energy of the trapped neutrinos 
$E^{\rm F}_{\nu_e}=p^{\rm F}_{\nu_e}=(6\pi^2n_{\nu_e})^{1/3} 
\sim 0.2 \rmmat{GeV}$, 
therefore, $E_{\nu}^{\rm tot}$ is given by,
\be
E_{\nu_e}^{\rm tot}\simeq\frac{3}{4}n_{\nu_e}E^{\rm F}_{\nu_e}\times 
\frac{4}{3}\pi R^3 
\simeq 2.12\times 10^{-2}
	\left(\frac{M}{2.0M_{\sun}}\right)^{4/3}
	\left(\frac{R}{1.5\times10^6 {\rm cm}}\right)^{-1}
	\left(\frac{Y_{\nu_e}}{0.1}\right)^{4/3} M_{\sun}.
\label{eq:neut}
\ee
It is evident that the total neutrino flux is of the same order for
the different initial masses.  The flux of the electron neutrinos in
both the NPH model and NP model is also similar, because the hyperons
only appear at a higher density when neutrinos are trapped, which makes
the hyperonic PNS and the normal PNS very similar.

\section{Conclusions and discussion}
\label{sec:conl}

In this paper, we investigate the rotational evolution of protoneutron
stars which contain only nucleons with that of PNSs which contain
hyperonic matter at zero temperature in full general relativity.  It
is found that PNSs contract and spin up at the early stage of its
evolution as the trapped neutrinos escape.  During this stage the
contribution of leptons to the pressure in the stellar crust
dominates. This conclusion is independent of the interior composition.
As the neutrinos continue to escape, the evolution of the PNS mainly
depends on the changing behavior of the EOS of the dense matter in the
stellar core; therefore, it differs in the two different models.  A
PNS which contains only nucleons stops contracting and begins to
expand because the EOS becomes stiffer as neutrinos escape from the
core.  PNSs that contain hyperonic matter keep shrinking, and their
spin frequency increases because the EOS becomes softer with the
escape of the neutrinos.  At the end of the evolution, for PNSs
without hyperons, the range of the spin periods becomes a little bit
narrower than the initial one.  However, the shape of the distribution
of the spin periods is very similar to that of the initial
distribution. For PNSs with hyperons, the distribution of the initial
spin periods is skewed significantly toward shorter periods.  If a PNS
is metastable, it keeps spinning up significantly on the neutrino
diffusion timescale before it collapses to a black hole.

In addition to the hyperonic matter, other exotic states, such as 
quark matter, kaon condensation and others might exist in the stellar
core. Because the neutrino trapping makes the onset of these exotic
states which contain negatively charged particles take place at the
higher baryon number density, a PNS with such an exotic composition
spins up as the neutrinos escape.  This characteristic phenomenon is
common among them.

In this work, the effect of the temperature is ignored.  Certainly
regardless of the interior composition, heat introduces another energy
source which would make the EOS stiffer than in the zero temperature
case.  The effect of the cooling of the PNS on the evolution of the
PNS could be predicted.  However, since the change of the chemical
equilibrium among the interior particles dominates the change of the
EOS, the cooling of the PNS could not change the results we obtain
qualitatively.

In principle, the rotation period of a PNS could be observed by a
ground-based gravitational radiation detector, such as LIGO. 
\citet{2002ApJ...565..430F} examine several avenues for gravitational
wave emission from core collapse.  If the core forms a bar or breaks
in two, the analysis that we have presented here is inapplicable
because we have assumed that the core remains axisymmetric
(Eq.~\ref{eq:axisym}).  Furthermore, the third mode that they
considered is the ringdown of the black hole; in this case, the
protoneutron-star stage is no longer visible.  However, they also have
estimated the role that $r-$modes may play in gravitational wave
generation \citep{2000ApJ...543..386H}.  The frequency of the observed
gravitational radiation reflects the spin frequency of the
protoneutron star, so the waveform would reveal the internal structure
of the protoneutron star as long as fallback did not strongly affect
the spin of the core during the Kelvin-Helmholtz epoch.  It is unclear
whether the $r-$modes will have a chance to grow during this early
epoch \citep{2002ApJ...565..430F}, but if they do LIGO-II could detect
gravitational radiation from supernovae within 10~Mpc.

As estimated by Eq.~\ref{eq:neut}, a huge energy
($10^{52}$~ergs) is released in the form of high energy neutrinos
during the Kelvin-Helmholtz epoch.  If the neutrino flux from the PNS
is modulated with the stellar rotation, a neutrino telescope might
observe temporal structure at the stellar period, which could
provide strict constraints on models of dense matter.
The typical energy of the emitted neutrinos is a few MeV
\citep{1989ARA&A..27..629A}.  Twenty neutrinos were detected from
SN1987A \citep{1987PhRvL..58.1490H,1987PhRvL..58.1494B} over about a
ten-second interval.  Currently, the total sensitivity of detectors
sensitive to supernova neutrinos is thirty times larger than in 1987,
so one would expect to detect about 15,000 neutrinos from a galactic
supernova.  \citet{2002AJ....124.1788R} give straightforward formulae
to estimate the minimum pulsed fraction detectable with a given
certainty from a sample of arrival times.  If we assume that the spin
frequency of the PNS lies between zero and 1500 Hz and that the spin
may vary by 200 Hz during the ten-second duration of the neutrino
burst, a blind search for the spin frequency and its evolution would
require about $3 \times 10^7$ trials.  If one wants to detect a signal
with a false alarm probability of $P_0$, the minimum sinusoidal pulsed
fraction that one could detect is
\begin{equation}
f_p = 2 \left [ \frac {\ln (N_\rmscr{trial}/P_0) - 1}{N} \right ]^{1/2} = 0.08
\end{equation}
for a false alarm probability of $10^{-3}$ and 15,000 detected
neutrinos.  Because the expected change in frequency is small compared
to the expected spin frequency, an unaccelerated search only does
marginally better with a minimum detectable pulsed fraction of 0.06.
Although these pulsed fractions are several times larger than required for
pulsar kicks
\citep[\protect{\em{e.g.}}][]{1998Natur.393..139S,1999ApJ...519..745A},
Earthbound neutrino detectors are most sensitive to neutrinos in the
Wien tail of the thermal distribution where the pulsed fraction will
be larger than the momemtum anisotropy.

A third observational probe of these models is the distribution of
initial periods of neutron stars.  This may be obtained from the
distribution of current spin periods of radio pulsars along with an
independent estimate of the ages of the neutron stars.  One must
assume that only magnetic dipole radiation has contributed to the spin
down, and furthermore that the standard model for magnetic dipole
radiation is correct \citep{2002nssr.conf....3K}. If gravitational
radiation is important, then such estimates of the initial spin
distribution probe the gravitational radiation epoch \citep[see][for a
few
viewpoints]{1998PhRvD..58h4020O,2000ApJ...543..386H,2002MNRAS.333..943W,Arras02}
rather than the protoneutron-star epoch discussed here.  Only a few
neutron stars have a good estimates of their initial periods, which
are typically an order of magnitude longer than the Kepler period.  If
the emission of gravititional radiation is not important to the spin
evolution of neutron stars, this would argue against a skewed
distribution of neutron-star initial spins; however, so few neutron
stars have estimates of their initial periods, and the contribution of
gravitational radiations is uncertain, it is premature to make any
definite conclusions.

The most spectacular manifestation of the rotational evolution of a
protoneutron star would be the formation and subsequent collapse of a
metastable stellar core.  How the core collapses into a black hole and
the observable consequences of that collapse are beyond the scope of
this paper.  We will examine the process in a subsequent paper.  For
a protoneutron star that rotates rapidly, an important issue is whether
the entire protoneutron star can collapse on a dynamical time or
portions of the star cannot fall directly into the black hole and form
an accretion disk which evolves on a viscous timescale.  This could
provide a detailed model of the central engines of collapsars, a
successful explanation of gamma-ray bursts
\citep{2001ApJ...550..410M}.

We have studied the structure of rapidly rotating axisymmetric
protoneutron stars using a fully general relativistic treatment of the
stellar structure and relativistic mean-field theory to model the
equation of state of dense matter.  We have contrasted the spin
evolution of normal protoneutron stars with hyperonic protoneutron
stars.  In general the evolution protoneutron stars with other exotic
species will be similar to that of hyperonic neutron stars.  We have
found that hyperonic protoneutron stars can be metastable; that is,
the maximum mass of a stable hyperonic protoneutron star decreases as
the neutrinos escape.  Both normal and hyperonic protoneutron stars
initially spin up as neutrinos escape the outer layers of the
protoneutron star and the low-density region shrinks.  As neutrinos
begin to escape the core of a hyperonic protoneutron star, the core
continues to spin up.  The metastable protoneutron star continues to
spin up as it begins to collapse to a black hole.  This contrasts with
behavior of a normal protoneutron star.  As neutrinos stream out the
core of a normal protoneutron star, the equation of state stiffens,
the core expands and spins down.  This spin evolution leaves a
signature on both the neutrino emission and gravitational radiation
from the stellar collapse that might be observable from nearby
supernovae.  Finally, if the processes leading to the formation of
hyperonic and normal protoneutron stars are similar, we would expect
that the hyperonic neutron stars initially to spin faster than normal
neutron stars.

\acknowledgments
%%%%%NEW ADDITION
We would like to thank the discussions with Prof. Ramesh Narayan,
Prof. K.S. Cheng, Prof. J.L. Zhang, Dr. Scott Ransom and Dr. John Bahcall.
Y.F.Y acknowledges the hospitality of Harvard-Smithsonian Center 
for Astrophysics.
%%%%%
Y.F.Y is partially supported by the Special Funds for Major State
Research Projects, and the National Natural Science Foundation
(10233030).  Support for J.S.H. was provided by the National
Aeronautics and Space Administration through Chandra Postdoctoral
Fellowship Award Number PF0-10015 issued by the Chandra X-ray
Observatory Center, which is operated by the Smithsonian Astrophysical
Observatory for and on behalf of NASA under contract NAS8-39073.

\bibliography{ms}

\end{document}